\documentstyle[12pt]{article}
\def\fakebold#1{\leavevmode\setbox0=\hbox{#1}%
  \kern-.025em\copy0 \kern-\wd0
  \kern .05em\copy0 \kern-\wd0
  \kern-.025em\raise.0433em\box0
}
\def\lapp{\mathrel{\mathop{<}\limits_{{}^\sim}}}
\def\bfg#1{\fakebold{$#1$}}
\def\sgb{\bfg{\sigma}}
\begin{document}
\vglue 2cm
\begin{center}
{\bf {\Large
         HADRON AND QUARK FORM FACTORS IN THE \\ 
         RELATIVISTIC HARMONIC OSCILLATOR MODEL \\}}
\vskip 2cm
{\large  { V. V. Burov }} \\
{\em  Laboratory  of  Theoretical  Physics,  Joint  Institute  for
Nuclear Research, \\
Head Post Office, Box 79, Moscow, U.S.S.R. } \\
\vspace{0.3cm}
{\large  A. De Pace } \\
{\em
Istituto Nazionale di Fisica Nucleare - sez. di Torino,\\
Via P. Giuria 1, I--10125, Torino, Italy.} \\
\vspace{0.3cm}
{\large  S. M. Dorkin } \\
{\em Far-East State University, Sukhanova 8,\\
Vladivostok, U.S.S.R.}\\
\vspace{0.3cm}
{\large  P. Saracco } \\
{\em Istituto Nazionale di Fisica Nucleare - sez. di Genova,\\
Via Dodecanneso 33, I--16146 Genova, Italy.}
\end{center}
\setlength{\baselineskip}{18pt}
\begin{abstract}
Nucleon, pion and quark form factors are studied within the relativistic 
harmonic oscillator model including the quark spin.
It is shown that the nucleon charge, magnetic and axial form factors and the
pion charge form factor can be explained with one oscillator parameter if one 
accounts for the scaling rule and the size of the constituent quarks. 
\end{abstract}

In this paper, we will analyze the proton electromagnetic and weak
form factors (FFs) and the pion charge FF in the framework of the 
Relativistic Harmonic Oscillator Model (RHOM).
Studies in this model were carried out in refs. \cite{Yukawa53}-\cite{BWS}.

In ref. \cite{Feyn71} a method was proposed
for building a covariant and gauge-invariant current within the
$ \tilde U(12)\otimes O(3,1) $ model.
There remain still open problems in that approach (see, for instance,
lectures \cite{Feyn72}). In refs. \cite{Blag74,Ishida77,Saito77} a covariant 
and gauge-invariant current
was found for the $ SU(6)\otimes O(3,1) $ model and it was shown that
all the nucleon FFs can be described in this case. 
However, the agreement with the experimental 
data is here much worse than within the nonrelativistic model that
takes account of the Lorentz
contraction of the nucleon wave function \cite{Beyer85}, although this latter
model fails to describe the electric FF of the neutron.
 
The aim of the present note is to describe the nucleon FFs within
RHOM using the $ SU(6)\otimes O(3,1) $ scheme of derivation of the covariant and
gauge-invariant currents under the assumption that the behavior of the FFs
when $ q^2 \to \infty $ is governed by the quark counting rules
\cite{Matveev} and that there holds the experimentally
observable scaling, i.e. $G_E^p(q^2)=G_M^p(q^2)/\mu _p = G_M^n(q^2)/\mu _n$, 
where $ \mu _{n,p} $ are the magnetic moments of neutron and proton.

Enforcing the scaling law leads one naturally to
introduce FFs for the constituent quarks. Actually, these could be
regarded as interpolating functions between the static quark model and the
asymptotic perturbative regime, since, as we shall see, they go to 1 both a
$q^2=0$ and $q^2\to\infty$.
 
Let us consider a system consisting of $N$ quarks in the field of a
relativistic harmonic oscillator. The corresponding wave function can be
represented in the form
\begin{eqnarray}
\Psi_p^{(N)}(x_1,\ldots, x_N) = \hat A\Phi_{N}
(x_1,x_2, \ldots x_N)U^{(N)}(\bf p \, ).
\label{WF}\end{eqnarray}
where $ \hat A $ is the operator of antisymmetrization of quarks
including the color  degrees of freedom to be not written for simplicity,
$ \Phi_{N}(x_1,x_2, \ldots x_N) $ is a covariant space -
time wave function, and $ U^{(N)}(\bf p \, ) $ is a spin wave function
to be described below. Let the wave function $\Phi_{N}$ obey the Klein-Gordon
equation with a relativistic harmonic oscillator potential
~\cite{Yukawa53} - ~\cite{BWS}
\begin{equation}
\left\{ \sum_{i=1}^{N} p_i^2 + {\cal K}^2\left[\sum_{i>j}^{N}\sum_{j=1}^{N-1}
(x_i-x_j)^2\right]\right\} \Phi_{N}(x_1,x_2, \ldots x_N) = 0,
\label{eq-RHOM}
\end{equation}
where $p_i=-i\partial/\partial x_i$ is a 4-momentum, ${\cal K}$ is the
oscillator parameter, $x_i$ is the 4-coordinate of the $ i $-th quark (we assume
all quark masses equal because of isospin invariance). Passing to
the center-of-mass coordinates $ X $ and the internal variables
$ r_0,\ldots r_{N-1} $ and diagonalizing, one can represent (\ref{eq-RHOM})
in the form
\begin{equation}
(p^2 -M_p^2) \Phi_{Nq}(r_0,r_1, \ldots r_{N-1},p) = 0, \quad
M_p^2 = -2 \alpha_N a_{i\mu}^+ a_{i\mu}+{\rm const}, \quad
\alpha_N= {\cal K} N\sqrt{N},
\label{alphaN}
\end{equation}
where
$ a_{i\mu } $ and $ a_{i\mu }^+ $ are, respectively, particle creation
and annihilation operators. Under the Takabayashi condition \cite{Takab65},
necessary for removing nonphysical oscillations along the coordinate of
relative time, $p^{\mu} a_{i\mu}^+ \Phi_{Nq} = 0,$
one gets the following solution
\begin{eqnarray}
\Phi_{Nq}(r_0,r_1, \ldots r_{N-1}, p)  & = &
\left(\frac{\alpha_N}{\pi N}\right)^{N-1} \exp (\frac{\alpha_N}{2N}
K^{\mu\nu} \sum_{i=1}^{N-1} r_{i\mu} r_{i\nu}), \\
\Phi_{N}(x_1,x_2, \ldots x_N)    & = & \exp \left\lbrack ip_\mu X_\mu
\right\rbrack
\Phi_{Nq}(r_0,r_1, \ldots r_{N-1}, p),
\label{wf-RHOM}
\end{eqnarray}
where $ p $ is the total momentum of the system and
$K^{\mu\nu} = g^{\mu\nu} - 2p^{\mu}p^{\nu}/p^2$.

The spin wave function can be constructed in two ways (see refs.
\cite{Ishida77,Honz85}) by transforming the nonrelativistic spin wave function
in the rest frame. The first method is to transform the wave function of
every quark separately as a Dirac spinor on the basis of the Bargmann-Wigner
equation. The second is to transform the wave function of the system as a
whole, the minimal Pauli transformation, since in this case the wave
function has a minimum number of components. The latter seems favorable,
as it allows for a non-identically-zero neutron charge FF,
in contrast to the first method. So, we will follow the second
method according to ref. \cite{Honz85}. Then, the spin wave function
$ U^{(N)}(\bf p \, ) $ can be represented in the form
\begin{eqnarray}
U^{(N)}(\bf p \,  )  & = & {B(\bf p \,  )}U^{(N)}(0), 
\qquad
U^{(N)}(0) = \left(
\begin{array}{c}
\chi \\ 0
\end{array}
\right),
\nonumber \\
\label{Uwf}\\
B(\bf p \,  ) & = & \exp\left\lbrack \frac{b}{2|{\bf p}|}\rho _1
(\bf p \cdot \sgb )
\right\rbrack  = \exp\left\lbrack \rho _1 b H\right\rbrack ,
\qquad\rho_1=\left(\begin{array}{cc}0&1\\1&0 \end{array}\right),
 \nonumber
\end{eqnarray}
where $\chi$ is the nonrelativistic spin function of the system,
$ H = {(\bf p \cdot \sgb )}/2|\bf p \, |$,  $ b= {\rm cosh}^{-1} p_0/M $ 
and $ \sgb \equiv \sum_{i=1}^N \sgb ^{i} $, $ \sgb ^{i} $ being
the Pauli matrices of the $ i $-th quark. 

Based on refs.\cite{Fujim71,Saito77,Honz85}, we write the
electromagnetic action in the form
\begin{equation}
I_{em}  =  \int\prod\limits_{i=1}^{N}dx_i\sum\limits_{k}^{N}
j_{k\mu }(x_1,\ldots, x_N)A_\mu (x_k) 
 \equiv  \int\!dX\,J_\mu^{(N)}(X)\,A_\mu(X), 
\label{Iem} 
\end{equation}
where
\begin{eqnarray}
j_{k\mu }(x_1,\ldots, x_N) = -i \bar \Psi_{p'}^{(N)} N e_k
\left\lbrack g_E(q^2)\frac{\stackrel\leftrightarrow\partial}
{\partial x_{k\mu }}
+i g_M(q^2)  \sigma _{\mu \nu }^{k}\left(\frac{\vec \partial }
{\partial x_{k\nu }}+
\frac{\stackrel\leftarrow\partial}{\partial x_{k\nu }}\right)\right\rbrack
\Psi_p^{(N)}.
\label{jmn}\end{eqnarray}
In eq.~(\ref{jmn}) $ \Psi_p^{(N)}$ ($\bar\Psi_{p'}^{(N)}$) is the initial 
(final) wave function of the $ N $ quark system (\ref{WF}), $ e_k $ is the 
charge of the $ k $-th quark, $  \sigma ^k_{\mu \nu } $ are the spin matrices 
of the $ k $-th quark 
($ \sigma _{ij}^k \equiv \varepsilon _{ijl}  \sigma _l^k$ ,
$ \sigma _{i4}^k= \sigma _{4i}^k \equiv \rho _1 \sigma _i^k$ ).
Assuming the constituent quarks to be not point particles, we identify
$ g_E(q^2) $ and $ g_M(q^2) $ with charge and magnetic quark FFs
($ q = p'-p $ is the 4-momentum transferred for an
 $ N $ quark system with an initial (final) 4-momentum $ p_\mu\ (p_\mu ') $).
Inserting the wave function (\ref{wf-RHOM}) into eqs.~(\ref{Iem}) and 
(\ref{jmn}) and computing the integrals over the internal quark variables 
$ r_0,\ldots, r_{N-1} $, one derives the matrix elements of the effective 
current for an N quark system between states of momentum $ p_\mu\ (p_\mu ') $ 
and spin component $ s\ (s') $:
\begin{eqnarray}
\left\langle p's'|J_\mu ^{(N)}(0)|ps\right\rangle =
\frac{I^{(N)}(q^2)}{\sqrt{2p_0p_0'}}\sum\limits_{k=1}^{N}
(\bar U_{s'}^{(N)}({\bf p'}) \Gamma_{k,\mu }U_s^{(N)}({\bf p} \, )),
\label{Ieff}\end{eqnarray}
where
\begin{eqnarray}
\Gamma_{k,\mu }=e_k[(p_\mu +p'_\mu )I_N(q^2)g_E(q^2)-iNg_M(q^2)
\sigma _{\mu \nu} ^{k}q_\nu ].
\label{Gamma}\end{eqnarray}
Here the overlapping integrals over space-time variables are the following
\begin{eqnarray}
I^{(N)}(q^2) & = & \frac{1}{\left(1+q^2/2M_N^2\right)^{(N-1)}}
\exp\left\lbrack -\frac{N-1}{4\alpha _N}\left(\frac{q^2}{1+q^2/2M_N^2}\right)
\right\rbrack \label{INq}\\
I_N(q^2)  & = & \frac{1+N q^2/2M_N^2}{1+q^2/2M_N^2},
\label{Int}\end{eqnarray}
$ M_N $ being the mass of the $N$-quark system. Using (\ref{Uwf}) 
one can write down eq.~(\ref{Ieff}) 
for a three-quark system, finally obtaining expressions for the nucleon FFs, 
namely
\begin{eqnarray}
F_M^{p,n}(q^2) &\equiv& \frac{G_M^p(q^2)}{\mu _p} = \frac{G_M^n(q^2)}{\mu _n} =
g_M(q^2) I^{(3)}(q^2),
\label{FM} \\
F_E^p(q^2) &\equiv& G_E^p(q^2) = \left\lbrack \left(1+\frac{q^2}{4M^2}\right)
I_3(q^2)g_E(q^2) - \frac{3q^2}{4M^2}g_M(q^2)\right\rbrack I^{(3)}(q^2),
\label{FE} \\
F_E^n(q^2) &\equiv& G^n_E(q^2)=\frac{q^2}{2M^2}g_M(q^2)I^{(3)}(q^2).
\label{FEn}\end{eqnarray}
 
It is also possible to compute the nucleon axial FF
$F_A(q^{2})=G_A(q^2)/G_A(0)$ (for details see ref. \cite{Saito77}),
\begin{eqnarray}
F_A(q^{2})=G_A(q^2)/G_A(0)=g_A(q^2)I_3(q^2)I^{(3)}(q^2),
\label{FA}\end{eqnarray}
where $g_A(q^2)$ is the constituent quark axial FF.
 
Note that within the $SU(6)$ model the magnetic moment of the proton equals
$ \mu _p=3 $, whereas for the neutron $ \mu _n=-2 $, somewhat different
from the experimental values $ \mu _p=2.793 $ and
$ \mu _n = -1.913 $.
Therefore, in the analysis of the experimental nucleon FFs
we will normalize them to 1 at $ q^2=0 $. In expressions (\ref{FM} - \ref{FA}) 
$ M\equiv M_3=0.938$ GeV is the nucleon mass.
 
Let us go back to the magnetic and charge nucleon FFs. 
According to the quark counting rules \cite{Matveev}
the nucleon FFs decrease as $ q^{-4} $ when $ q^2 \to\infty $.
From inspection of formulae (\ref{INq}-\ref{FA}), it is clear that all the FFs
calculated in this model have the correct asymptotic behaviour already
with point-like constituent quarks. For instance, when $ q^2\to\infty $, 
$ F_M^p(q^2) \sim g_M(q^2)/q^4$, therefore it is natural to make the minimal 
assumption that $ g_M(q^2)=1 $. However, since from the scaling condition it 
also follows that $ F_M^p(q^2)=F_E^p(q^2) $, we can easily derive the following 
expression for the quark charge FF $ g_E(q^2) $:
\begin{eqnarray}
g_E(q^2)=\frac{\left(1+3q^2/4M^2\right)\left(1+q^2/2M^2\right)}
{\left(1+q^2/4M^2\right)\left(1+3q^2/2M^2\right)} g_M(q^2).
\label{GE}\end{eqnarray}
We stress that eq.~(\ref{GE}) is dictated by the scaling condition,
which would be violated for point-like quarks, and not by the asymptotic
dependence on $q$ of the FFs, which comes out naturally from the model.
 
The only free parameter in this model is the oscillator parameter: fitting 
(\ref{FM}) to the experimental data for the proton FFs, 
we find $ \alpha _3  = 0.42$ (GeV/c)$^2$, in agreement with the results
of ref. \cite{Saito77}.
From formula (\ref{FE}) one determines the theoretical value of the slope
of the electric FF at $ q^2=0 $ to be
$-d F_E^n(q^2)/d q^2|_{q^2 = 0}= 0.022$, which is consistent with the known
experimental value $0.0202 \pm 0.0003$ fm$^2$ \cite{neutron}.

The standard parametrization of the nucleon FFs is in terms of a dipole
function
\begin{eqnarray}
F_D(q^2)=\frac{1}{\left(1+q^2/\Lambda^2_{E,A}\right)^2},
\label{dipol}\end{eqnarray}
where for the nucleon electromagnetic FFs
$ \Lambda^2_E=0.71$ (GeV/c)$^2 $ and
for the axial nucleon FF $\Lambda_A=(1.032\pm0.036)$ GeV/c
\cite{Miller}. 

In Figs.~1 we show the comparison of our computations within RHOM
with the experimental data for the nucleon electromagnetic FFs divided 
(except for $F^n_E$) by the dipole parametrization (\ref{dipol}).
Note that the proton FFs have
been fitted (adjusting $\alpha_3$) to experiment, whereas the others have been 
found without additional parameters. 
In the case of the neutron, the standard Galster parametrization \cite{Galster}
is also shown: actually, by using an effective mass of $\sim 1$ GeV in the RHOM
calculation one can closely reproduce the Galster curve.

In Fig.~2 the dipole parametrization (solid lines) of the axial FF is 
compared to formula (17) with $g_A(q^2)=1$ (dotted line) and 
$g_A(q^2)=g_E(q^2)$ (dashed line). Note the improved agreement in the latter
case: the effect of the constituent quark size, although not dramatic, is not
negligible ($\sim10\%\div15\%$).
 
It is straightforward to estimate the nucleon and quark sizes from  equations
(\ref{FM}), (\ref{FE}) and (\ref{GE}):
$\bar r _{(p,n)}=\sqrt{<r^2_{(p,n)}>} = \sqrt{6(1/2\alpha _3 + 1/M^2)}$,
$\bar r _{q}=\sqrt{<r^2_{q}>} = \sqrt{3/M^2}$. Thus,
$\bar r_{(p,n)}=0.74$ fm ($\bar r^{\it exp}_{(p,n)}\cong(0.87\pm0.07)$ fm),
and $\bar r_q=0.36$ fm.
From (\ref{FEn}), on the other hand, one finds 
$<r^2>_{E,n}=-3/M^2=-0.13$ fm$^2$ ($<r^2>^{\it exp}_{E,n}=(-0.119\pm0.004)$ 
fm$^2$).

Concerning the axial radius, it is interesting to note that in the $g_A=1$ case
$\sqrt{<r^2>_A}=\sqrt{3/\alpha_3}=0.53$ fm, whereas for $g_A(q^2)=g_E(q^2)$
$\sqrt{<r^2>_A}=\sqrt{3(1/\alpha_3+1/M^2)}=0.64$ fm (to be compared with the
experimental value of ($0.65\pm0.07)$ fm).
 
The procedure outlined above can also be applied to the quark-antiquark system.
Here a source of ambiguity is due to the two possible representations for the
booster in eq.~(\ref{Uwf}): indeed, both $\sgb=\sgb^1+{\overline{\sgb}}^2$
and $\sgb=\sgb^1-{\overline{\sgb}}^2$ lead to a valid formulation
\cite{Ishida77}. However, the latter case is known to give a simple pole-like
asymptotic behaviour for the pion FF only for a special combination of
the single-quark currents in eq.~(\ref{jmn}) (see ref.~\cite{Ishida79}).
In the following we shall stick to the first formulation, where the pion charge
FF reads
\begin{equation}
  F_\pi(q^2) = g_E(q^2) I_2(q^2) I^{(2)}(q^2), 
\label{Fpion}
\end{equation}
$g_E(q^2)$ being still given by eq.~(\ref{GE}). 

Assuming the same oscillator parameter ${\cal K}$ for the two- and three- quark 
systems, the value of $\alpha_2$ can be derived from $\alpha_3$ since, using 
eq.~(\ref{alphaN}), $\alpha_2/\alpha_3=(2/3)^{(3/2)}$. Then, from 
$\alpha_3=0.42$ (GeV/c)$^2$, one gets $\alpha_2=0.23$ (GeV/c)$^2$.
Note that in the literature both $\alpha_2$ and $\alpha_3$ have usually been 
regarded as free parameters (see, e.~g., \cite{Ishida79,Kizu79}).
A known problem is connected to the mass of the quark-antiquark system: indeed,
using the physical pion mass one largely underestimates $F_\pi(q^2)$.
In the literature a wide range of ``$SU(6)$ symmetric masses'' has been employed
($0.5$ GeV$\lapp M_2\lapp 0.8$ GeV) \cite{Ishida77,Saito77,Ishida79,Kizu79} and
in the following we use $M_2=0.77$ GeV. However, it is found that this is not
a critical parameter, since expression (\ref{Fpion}) is mildly dependent on
$M_2$ for $0.4$ GeV $\lapp M_2\lapp 1$ GeV, giving in this range descriptions of
the data of comparable quality (note that the same is not true for the case with
$g_E=1$, which is much more sensitive to $M_2$).

In Fig.~3 we compare with the experiment our calculations of $F_\pi(q^2)$ using
either $g_E(q^2)$ given by eq.~(\ref{GE}) and $g_E(q^2)=1$. 
The improvement due to the constituent quark FF is apparent.
It is again interesting to note that the pion radius is independent of $M_2$ and
turns out to be 
$\bar r_\pi=\sqrt{<r^2>_\pi}=\sqrt{3(1/2\alpha_2+1/M^2)}=0.62$ fm and 
$\bar r_\pi=\sqrt{3/2\alpha_2}=0.50$ fm for $g_E(q^2)$ given by eq.~(\ref{GE})
and $g_E=1$, respectively ($\bar r^{\it exp}_\pi=(0.66\pm0.01)$ fm).

We summarize here our conclusions.
This model gives a simple description of the experimental data on nucleon and 
pion FFs provided only one arbitrary parameter is used. 
Note that the quality of the description can be improved upon by using 
$g_M(q^2)$ and $g_A(q^2)$ to fit the nucleon FFs.
The model could then be used, with all the parameters fixed, in investigations,
for instance, of the strong $\pi$NN vertex, 
$ \Delta $ - isobar FFs, nucleon structure functions and so on.  
Besides, the model can be easily extended to systems with a number of quarks 
larger than three, e.~g. to the study of the deuteron FFs \cite{Kizu79,
Honz85,BWS}.

\newpage
\vspace*{3cm}
 
\centerline{\bf Figure captions  }
\bigskip \bigskip 
\begin{itemize}
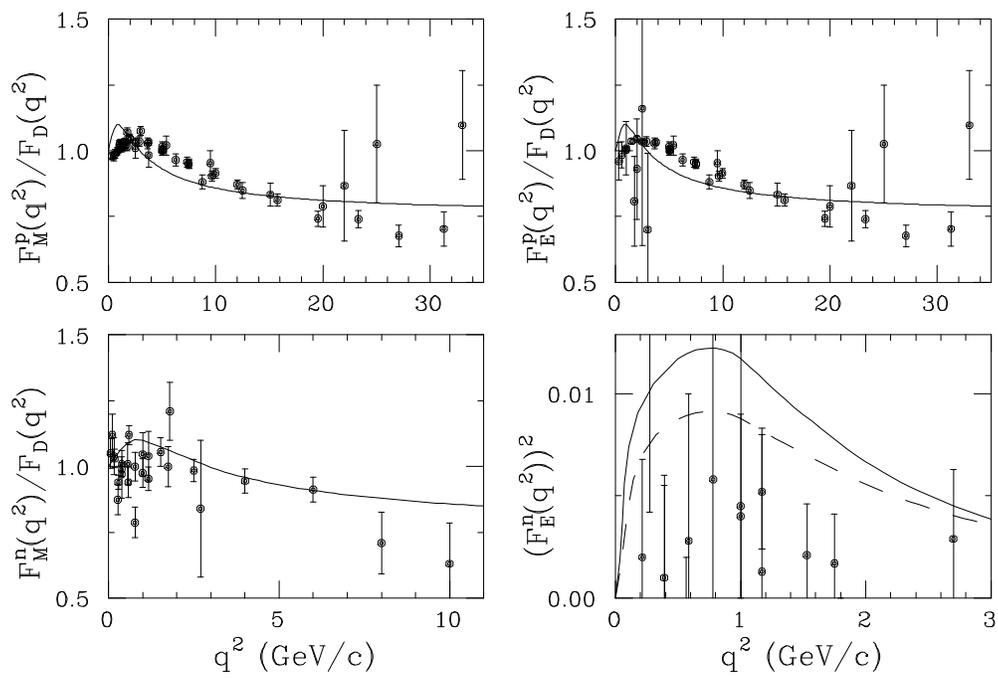
\item [Fig.1]
The solid line is the ratio of the normalized electromagnetic FFs of 
the nucleon to the dipole parametrization (\protect\ref{dipol});
in the case of the neutron the squared electric FF is shown (solid
line), compared to the Galster parametrization $F^n_E=-\mu_n(q^2/4M^2)F_D$
(dashed line). The experimental  data  are taken from 
refs.~\cite{Bartel}.
\bigskip 

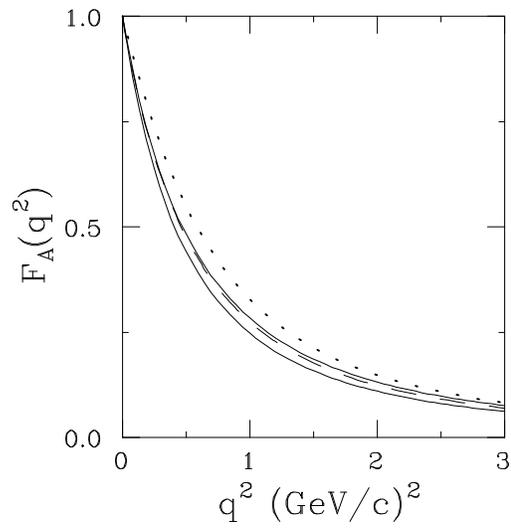
\item [Fig.2]
The normalized axial FF of the nucleon  $F_A(q^{2})=G_A(q^2)/G(0)$.
The solid lines represent the dipole fits of the neutrino experiments and
correspond to the upper and lower world average values for $\lambda_A$
\protect\cite{Miller}; they are compared to 
the present calculation in the framework of RHOM with $g_A(q^2)=1$ (dotted line)
and $g_A(q^2)=g_E(q^2)$ (dashed line).
\bigskip 

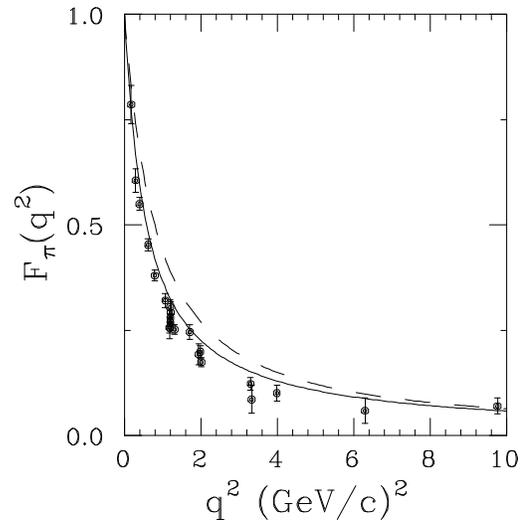
\item [Fig.3]
The pion charge FF (\ref{Fpion}) with $g_E(q^2)$ either given by 
eq.~(\ref{GE}) (solid line) or equal to 1 (dashed line). 
The experimental  data  are taken from refs.~\cite{Bebek}.
\end{itemize}
\end{document}